\documentclass[onecolumn,showpacs,aps,amssymb,floatfix,prd,amsmath,preprintnumbers]{revtex4}
\usepackage{epstopdf}
\usepackage{capt-of}
\usepackage{graphicx}  
\usepackage{dcolumn}   
\usepackage{float}
\usepackage{hyperref}
\begin{document}
\title{Existence of bulk viscous universe in $f(R,T)$ gravity and confrontation with observational data}

\author{Anil Kumar Yadav$^{1}$\footnote{Email: abanilyadav@yahoo.co.in},
Lokesh Kumar Sharma$^{2}$\footnote{Email: lksharma177@gmail.com},    
B. K. Singh$^{2}$\footnote{Email:  benoy.singh@gla.ac.in},
P. K. Sahoo$^{3}$\footnote{Email: pksahoo@hyderabad.bits-pilani.ac.in}}\
  
\affiliation{$^{1}$Department of Physics, United College of Engineering and Research, Greater Noida - 201306, India}
\affiliation{$^{2}$Department of Physics, GLA University, Mathura - 281406 India}
\affiliation{$^{3}$ Department of Mathematics, Birla Institute of
Technology and Science-Pilani, Hyderabad Campus, Hyderabad-500078,
India}

\begin{abstract}

\textbf{Abstract:} In this paper we have investigated a bulk viscous universe in $f(R,T)$ gravity where $R$ and $T$ are the Ricci scalar and trace of energy momentum tensor respectively. We have obtained explicit solutions of field equations in modified gravity by considering the power law form of scale factor. The Hubble parameter and deceleration
parameter are derived in terms of cosmic time and redshift both. We have estimated the present values of these parameters with observational Hubble data and SN Ia data sets. At 1$\sigma$ level, the estimated values of $q_{0}$ and $m$ are obtained as $q_{0}=-0.30 \pm 0.05$ \& $ m = 0.70 \pm 0.02 $ where $q_{0}$ is the present value of deceleration parameter and $m$ is the model parameter. The energy conditions and Om(z) analysis for the anisotropic LRS Bianchi type I model are also discussed.

\end{abstract}


\maketitle

\section{Introduction}\label{1}

From the recent astrophysical observations: $H(z)$ from Ia supernova \cite{Perlmutter/1998,Perlmutter/1999,Riess/1998,Tonry/2003,Clocchiatti/2006}, CMB \cite{Bernardis/2000,Hanany/2000}, baryon acoustic oscillations (BAO) \cite{Blake/2011,Padmanabhan/2012,Anderson/2013} and PLANK \cite{Bennett/2013}, it has been confirmed that we are lived in an accelerating universe. The main conclusion of these observations are strongly suggest that nearly two-third of the total energy density is in the form of unknown/mysterious energy. This late time acceleration of the universe is considered to be driven dark energy (DE) however the actual nature of DE is yet to be investigate. In the literature, numerous models with different DE candidates are proposed. Among these models, $\Lambda$CDM model is more acceptable by theoretical physicists to explain the behavior of DE and late-time acceleration of universe. The $\Lambda$CDM models have two fundamental problems - fine-tuning at plank scale and cosmic coincidence \cite{Nima/2000}. The problems associated with $\Lambda$CDM model have led to search for the reconstruction of gravitational field theories which could be capable of reproducing late time cosmic acceleration without inclusion of cosmological constant in Einstein's field equation. However one can not claim that the idea of modification in general relativity was en-lighted just after the discovery of accelerating universe \cite{Brans/1961,Lyra/1951}. A number of modified theories such as $f(R)$ theory \cite{Carroll/2004,Nojiri/2007}, $f(G)$ gravity \cite{Nojiri/2005}, $f(T)$ theory \cite{Linder/2010,Myrzakulov/2011} and $f(R,T)$ theory \cite{Harko/2011} exist since a long time due to the combined need of astrophysics, high energy physics and cosmology. At beginning, the quest in modification of general relativity was focused on the change of geometrical part of Einstein-Hilbert gravitational action. In \cite{Harko/2011} the authors have investigated a non-minimal coupling between matter and geometry in the framework of an effective gravitational Lagrangian consisting of $R \ \ \& \ \ T$: $R$ is an arbitrary function of Ricci scalar and $T$ is the trace of energy-momentum tensor and thus introduced $f(R,T)$ theory of gravitation. In this theory, the choice of $T$ is due to the existence of some imperfect fluids. Thus, the $f(R,T)$ theory of gravitation may give a complete description of late time acceleration of universe without resorting the existence of dark energy. This extraordinary features of $f(R,T)$ theory of gravitation has attracted researchers to study and reconstruct this theory in various contexts of astrophysics and cosmology \cite{Nojiri/2011,Shabani/2013}. Recently, Nojiri et al. \cite{Nojiri/2017} have studied inflation, bounce and late time evolution in modified theory of gravity. It is important to mentioning that $f(R,T)$ theory is also applicable to investigate the effects of expansion-free condition in the formulation of structure scalars. Some useful applications and existence of  relativistic stellar objects in $f(R,T)$ theory of gravity are given in references \cite{Yousaf/2016,Yousaf/2019,Yousaf/2017}.\\

After discovery of Wilkinson Microwave Probe \cite{Jaffe/2005,Hinshaw/2009}, the homogeneous and anisotropic models have been gaining an increasing attention and tremendous momentum in observational cosmology in the search of relativistic picture of the universe in its early stages. A spatially homogeneous Bianchi I (BI) space-time necessarily have three dimensional group, which acts as transitively on space-like three dimensional orbits. Therefore, the universe should achieve following two features: (i) a slightly anisotropic geometry in spite of inflation, and (ii) a non trivial isotropization history of the universe due to the presence of an anisotropic energy source. The advantage of these anisotropic model are that they have a significant role in the description of evolution of early phase of the universe and capable in finding more general cosmological models in comparison to FRW model. The LRS BI universe is homogeneous and anisotropic model of universe and it is differ from FRW model: a popular model to describes the fate of universe. But, in the recent time, CMB observation indicates very tiny variations in the intensities of microwaves coming from different direction in the sky \cite{Amirhashchi/2018}. This observation challenges the isotropic assumptions in spatial directions so the idea of anisotropic space-time comes forward with Bianchi type models \cite{Yadav/2012,Goswami/2015,Goswami/2016,Sharma/2019,Yadav/2011IJTP,Yadav/2011ASTR,Yadav/2016,Yadav/2019,Kumar/2011}. Recently, in \cite{Mishra/2018,Mishra/2018a} the authors have investigated some non-isotropic cosmological models in different physical contexts.

In the realm of cosmology, especially bulk viscous phenomena have attracted considerable interest \cite{Graon/1990} because it is only possible dissipative mechanism in Bianchi type space-time \cite{Misner/1968,Zimdahl/2001}. The coefficient of bulk viscosity vanishes both for actual relativistic and non-relativistic equation of state. In the era of inflationary phase, the contribution of bulk viscosity is well recognized. The basic concept of bulk viscous driven inflation is that the bulk viscosity contributes a negative pressure and this negative pressure simulating a repulsive gravity of the matter and gives an impetus for rapid expansion of the universe. Relativistic cosmological solutions and the different phases of the universe with non-causal viscous fluid are studied in $f(R,T)$ gravity theory \cite{Singh/2014}. In this extended theory of gravity, ideal fluids play major role to contribute acceleration but on the hydrodynamics scale, it is impossible to define turbulence phenomena without inclusion of bulk viscosity. The two viscous coefficients are discussed in literature: (i) shear viscosity and (ii) bulk viscosity. In early universe, Hogeveen and his team have examined the contribution of both shear and bulk viscosity by using kinetic theory and it is discovered that the impact of viscosity is very small in early universe whereas the impact may be significant in future universe \cite{Hogeveen/1986}. Later on, Brevik and his co-authors have investigated the effect of viscosity in early universe for both homogeneous and in-homogeneous equation of state \cite{Brevik/2017,Brevik/2017a}. Recently, in  \cite{Mishra/2019,Yadav/2019mpla,sahoo/2017a} the authors have investigated bulk viscous embedded hybrid universe in general relativity and $f(R,T)$ theory of gravitation respectively.\\
  
The main goal of this paper is to investigate the bulk viscous anisotropic universe in the framework $f(R,T)$ theory of gravity. The paper is organized as follows: Section \ref{section2} represents the explicit solutions of field equations of LRS BI universe. Section \ref{section3} deals with the confrontation of derived model with observational data. The behaviour of energy conditions are discussed in Section \ref{section4}. In Section \ref{section5}, we present the Om(z) analysis and finally the conclusion is given in section \ref{section6}.

\section{Metric and field equations}\label{section2}
The spatial homogeneous and anisotropic LRS BI metric is read as
\begin{equation}
\label{metric}
ds^{2} = -dt^{2} + A^{2}(t) dx^2+ B^{2}(t)(dy^{2}+dz^{2}) 
\end{equation} 
where $A(t)$ and $B(t)$ are scale factors along the spatial direction.\\
The energy momentum tensor for bulk-viscous fluid is read as
\begin{equation}
\label{eq2}
T_{\mu \nu} = (\rho+\bar{p})u_{\mu}u_{\nu}-\bar{p}g_{\mu\nu}
\end{equation}
where $u_{\mu} = (0,0,0,1)$ is the four velocity vector in co-moving co-ordinate system satisfying $u_{\mu}u_{\nu} = 1$ and $\bar{p} = p - 3\xi H$.\\
Here, $\xi$, $\bar{p}$ and $p$ are the bulk viscous coefficient, bulk viscous pressure and normal pressure respectively.\\
The energy-momentum tensor for barotropic bulk viscous fluid under $f(R,T) = R + 2\zeta T$ formalism is given by
\begin{equation}
\label{eq3}
R_{\mu\nu}-\frac{1}{2}Rg_{\mu\nu} = 8\pi T_{\mu\nu}-2(T_{\mu\nu}+\Theta_{\mu\nu})f^{\prime}(T)+f(T)g_{ij}
\end{equation}
where $f(T) = \zeta T$, $\zeta$ is an arbitrary constants.\\
Thus, the equation (\ref{eq3}) and (\ref{metric})lead to
\begin{equation}
\label{fe4}
-2\frac{\ddot{B}}{B}-\frac{\dot{B}^{2}}{B^{2}} = (8\pi+3\zeta)\bar{p}-\zeta \rho
\end{equation}
\begin{equation}
\label{fe5}
-\frac{\ddot{A}}{A} -\frac{\ddot{B}}{B}-\frac{\dot{A}\dot{B}}{AB} = (8\pi+3\zeta)\bar{p}-\zeta \rho
\end{equation}
\begin{equation}
\label{fe6}
2\frac{\dot{A}\dot{B}}{AB}+\frac{\dot{B}^{2}}{B} = (8\pi+3\zeta)\rho-\zeta \bar{p}
\end{equation}
Equation (\ref{fe4}) -(\ref{fe6}) are the system of three equations with four unknown variables. In general, it is impossible to solve these equations however the exact solution is only possible by taking into account at least one physical relation among parameters. So, we assume that the average scale factor expands in power function of time i.e.\\
\begin{equation}
\label{a}
a = (mDt)^{\frac{1}{m}}
\end{equation}
where $m$, $D$ and $n$ are positive \textbf{constants}.\\
Equation (\ref{a}) gives the following expressions for directional scale factors $A$ and $B$
\begin{equation}
\label{eq7}
A = (m_{1}D_{1}t)^{\frac{1}{m_{1}}}
\end{equation}
\begin{equation}
\label{B}
B = (m_{2}D_{2}t)^{\frac{1}{m_{2}}}
\end{equation}
where $m_{1}$, $m_{2}$, $D_{1}$ and $D_{2}$ are constant and satisfies the following relation.\\
$m(m_{1}+2m_{2}) = 3m_{1}m_{2}$, $D^{\frac{3}{m}} = D_{1}^{\frac{1}{m_{1}}}D_{2}^{\frac{2}{m_{2}}}$\\
  
The Hubble's parameter $(H)$ and deceleration parameter $(q)$ are given by
\begin{equation}
\label{eq11}
H = \frac{1}{mt}
\end{equation}
\begin{equation}
\label{q}
q = m-1
\end{equation}
The energy density $(\rho)$, pressure (p) and bulk viscous pressure $(\bar{p})$ for model (\ref{metric}) are read as
\begin{equation}
\label{eq13}
\rho = \frac{1}{8\pi+\zeta(3-\gamma)}\left[\frac{2m_{2}+m_{1}}{m_{1}m_{2}^{2}t^{2}}-\frac{3\xi}{m m_{2}^{2}t^{3}}\right]
\end{equation}
\begin{equation}
\label{eq14}
p = \frac{\gamma}{8\pi+\zeta(3-\gamma)}\left[\frac{2m_{2}+m_{1}}{m_{1}m_{2}^{2}t^{2}}-\frac{3\xi}{m m_{2}^{2}t^{3}}\right]
\end{equation}
\begin{equation}\label{eq15}
\bar{p} = \frac{\gamma}{8\pi+\zeta(3-\gamma)}\left[\frac{2m_{2}+m_{1}}{m_{1}m_{2}^{2}t^{2}}-\frac{3\xi}{m m_{2}^{2}t^{3}}\right]-\frac{3\xi}{mt}
\end{equation}

\section{Confrontation with observational data}\label{section3}

The relation between scale factor $a$ and redshift $z$ is given by

\begin{equation}
\label{z}
a=\frac{a_{0}}{1+z}
\end{equation}
where $a_{0}$ is the present value of scale factor.\\
The differential age of the galaxies are used for observational Hubble data (OHD) by following equation \cite{Akarsu/2014}
\begin{equation}
\label{z-3}
H(z) = -\frac{1}{1+z}\frac{dz}{dt}
\end{equation}
We consider 36 data points of OHD in the redshift range $0.07\leq z \leq 2.36$ with their corresponding standard deviation $\sigma_{H}$, recently compiled in Refs. \cite{Amirhashchi/2018,Akarsu/2019}.\\

We define matter energy density parameter $(\Omega_{m})$, anisotropy curvature parameter $(\Omega_{\sigma})$ and bulk viscous parameter $(\Omega_{b})$ as following
\begin{equation}\label{density}
\Omega_{m} = \frac{8\pi\rho}{3H^{2}},\;\;\Omega_{\sigma} = \frac{\sigma^{2}}{3H^{2}a^{6}},\;\;\Omega_{b} = \frac{\xi_{1}}{3H^{2}}
\end{equation}
and
\begin{equation}
\label{flat}
\Omega_{m}+\Omega_{\sigma}+\Omega_{b} = 1
\end{equation}
where $\sigma^{2} = \frac{1}{2}\sigma_{ij}\sigma^{ij}$ is the shear scalar and $\xi_{1} = 3\zeta(\rho+\xi H)$.\\
 
Solving equations (\ref{fe6}), (\ref{z}), (\ref{density}) and (\ref{flat}), we obtain
\begin{equation}
\label{H}
H^{2} = H_{0}^{2}[(\Omega_{m})_{0}(1+z)^{3}+(\Omega_{\sigma})_{0}(1+z)^{6}+(\Omega_{b})_{0}]
\end{equation}
where subscript 0 denotes the present value of parameters. $H_{0}$ is the present value of Hubble constant in $km s^{-1}Mpc^{-1}$. Here we find constraints on $H_{0}$ and $(\Omega_{b})_{0}$ by using 36 H(z) data points, recently published in Table II of the paper of Akarsu et al \cite{Akarsu/2019}. For this sake, we define $\chi^{2}$ as following
\begin{equation}
\label{chi1}
\chi^{2}_{OHD}(H_{0},(\Omega_{b})_{0}) = \sum_{i=1}^{36}\left[\frac{H(z_{i},H_{0},(\Omega_{b})_{0})-H_{obs}(z_{i})}{\sigma_{i}}\right]^{2}
\end{equation}
where $H_{obs}(z_{i})$ is the observed value of Hubble parameter with standard deviation $\sigma_{i}$ and 
$H(z_{i},H_{0},(\Omega_{b})_{0})$ is the theoretical values obtained from the derived model. We find that the best fit values of the parameters are $H_{0} = 65.53 \pm 1.9$ \& $ (\Omega_{b})_0 = 0.68 \pm 0.06 $ together with reduced $\chi^{2}_{OHD} = 7.61$. The likelihood contour at 68.3\% , 95.4\%  and 99.7\% confidence level around the best fit values in the $H_{0} - (\Omega_{b})_0$ plane is shown in Fig. 1. Similarly Fig. 2 depicts the likelihood contour at 68.3\%, 95.4\% and 99.7\%  confidence level around the best fit values as $q_{0}=-0.32 \pm 0.03$ \& $m = 0.68 \pm 0.04 $ in the $m - q_{0}$ plane obtained by fitting derived model with 36 observation Hubble data points.\\
\begin{figure}[H]
\begin{center}
\includegraphics[width=8.0cm]{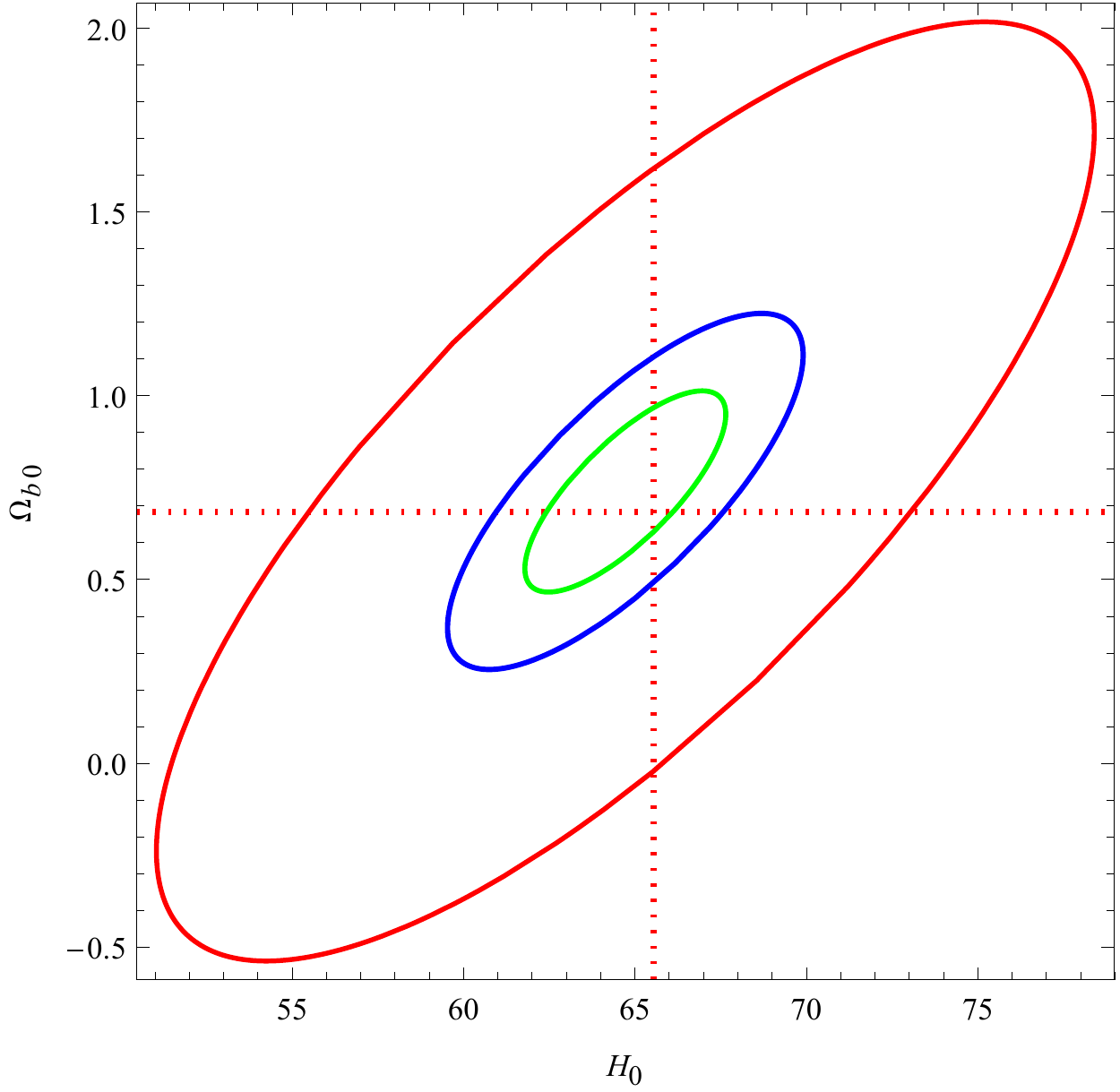}
\end{center}
\caption{The likelihood contour at 68.3\% (inner contour), 95.4\% (middle contour) and 99.7\% (outer contour) confidence level around the best fit values as $H_{0}=65.53 \pm 1.9$ \& $ (\Omega_{b})_0 = 0.68 \pm 0.06 $ in the $H_{0} - (\Omega_{b})_0$ plane obtained by fitting derived model with $H(z)$.}
\end{figure}
\begin{figure}[H]
\begin{center}
\includegraphics[width=8.0cm]{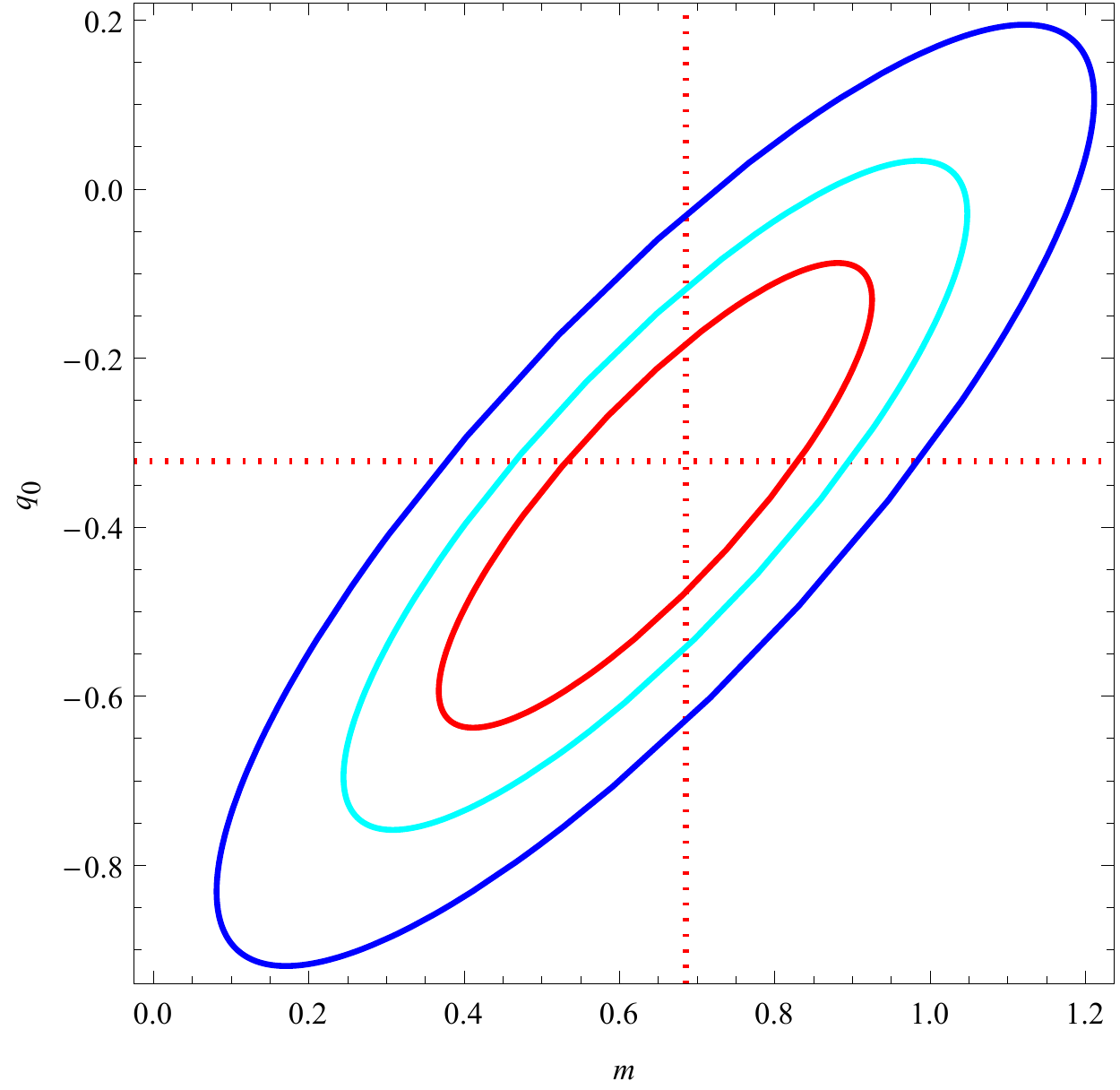}
\end{center}
\caption{The likelihood contour at 68.3\% (inner contour), 95.4\% (middle contour) and 99.7\% (outer contour) confidence level around the best fit values as $q_{0}=-0.32 \pm 0.03$ \& $m = 0.68 \pm 0.04 $ in the $m - q_{0}$ plane obtained by fitting derived model with 36 observation Hubble data points.}
\end{figure}
\subsection{Type Ia Supernova}
We fit our model with 580 points of SN Ia data set \cite{Suzuki/2012} and choose the value of current Hubble constant as $H_{0} = 65.53~kms^{-1}Mpc^{-1} $  to complete the data set. Thus $\chi^{2}_{SN}$ is obtained as
\begin{equation}
\label{chi2}
\chi^{2}_{SN}(m, q_{0}) = \sum_{i=1}^{580}\left[\frac{\mu(z_{i},m, q_{0})-\mu_{obs}(z_{i})}{\sigma_{\mu}(z_i)}\right]^{2}
\end{equation}
where $\mu(z_{i},m, q_{0})$ and $\mu_{obs}(z_{i})$ are the theoretical and observed values distance modulus for the model under consideration respectively. $\sigma_{\mu}(z_i)$ represents the standard error in the observed value of $\mu$. The distance modulus is given by
\begin{equation}
\label{dm}
\mu(z) = m_{b}-M = 5log_{10}D_{L}(z) + \mu_{0}
\end{equation} 
where $m_{b}$, M are the apparent magnitude and absolute magnitude of a standard candle respectively. The luminosity distance $D_{L}$ and nuisance parameter $(\mu_{0})$ are read as
\begin{equation}
\label{dl}
D_{L} = \frac{c(1+z)}{H_{0}}\int_{0}^{z}\frac{dz}{h(z)};\;\;\; h(z)=\frac{H}{H_{0}}
\end{equation} 
and
\begin{equation}
\label{nu}
\mu_{0} = 5log_{10}\left(\frac{H_{0}^{-1}}{Mpc}\right) + 25
\end{equation} 
Therefore, the distance modulus and apparent magnitude are given by
\begin{equation}
\label{mu-2}
\mu(z) = 5log_{10}\left[\frac{c(1+z)}{H_{0}}\int_{0}^{z}\frac{dz}{h(z)}\right]
\end{equation}
\begin{equation}
\label{mb-1}
m_{b}(z) = 16.08+5log_{10}\left[\frac{1+z}{0.026}\int_{0}^{z}\frac{dz}{h(z)}\right]
\end{equation}
Thus $\chi_{total}^{2}$ is given by
\begin{equation}
\label{chi3}
\chi_{joint}^{2} = \chi_{OHD}^{2} + \chi_{SN}^{2} 
\end{equation}
Fig. 3 shows the likelihood contour at 68.3\%, 95.4\% and 99.7\% confidence level around the best fit values as $q_{0}=-0.30 \pm 0.05$ \& $ m = 0.70 \pm 0.02 $ in the $m - q_{0}$ plane obtained by fitting derived model with $H(z)$ + SN Ia data.\\
\begin{figure}[H]
\begin{center}
\includegraphics[width=8.0cm]{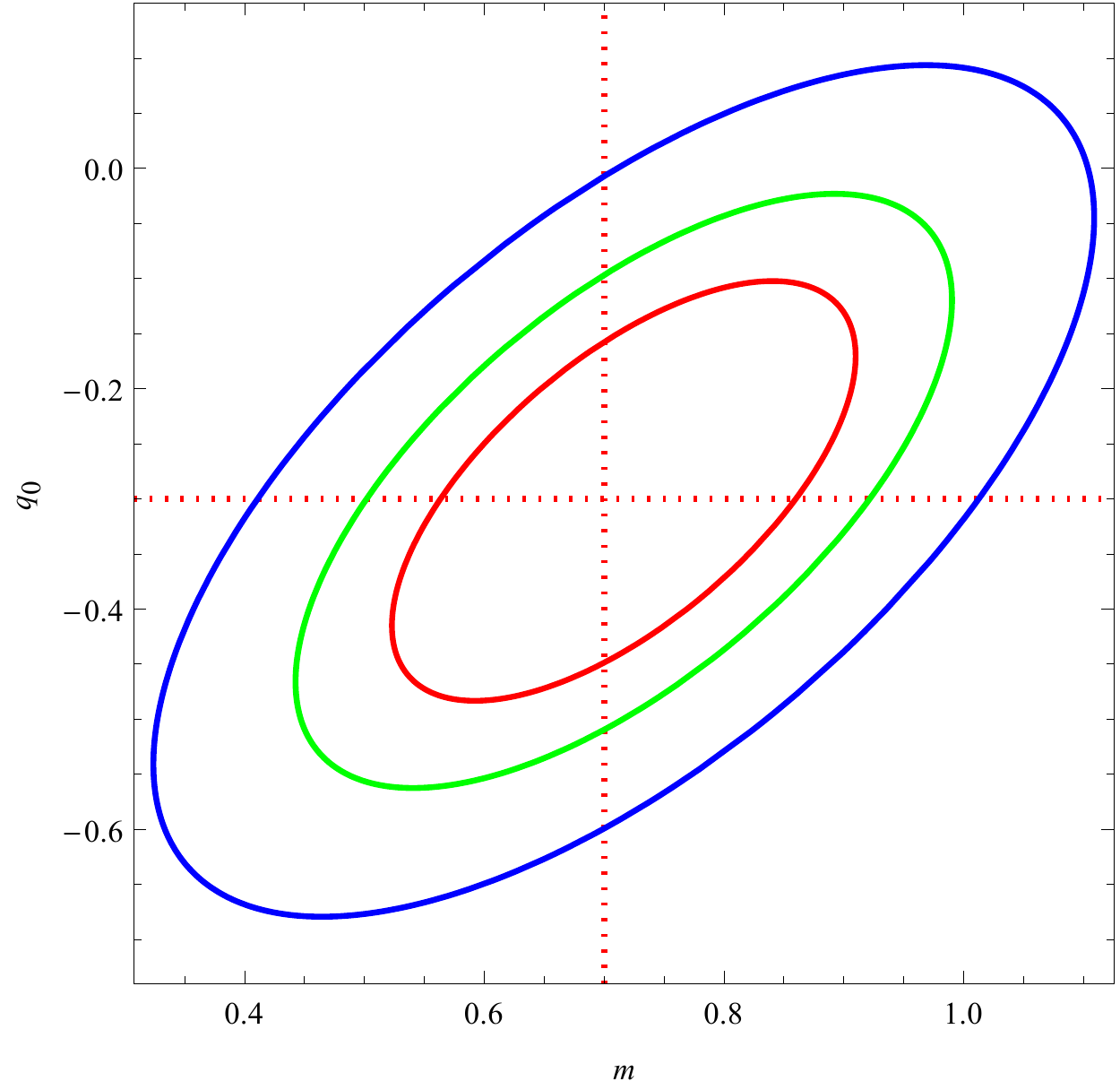}
\end{center}
\caption{The likelihood contour at 68.3\% (inner contour), 95.4\% (middle contour) and 99.7\% (outer contour) confidence level around the best fit values as $q_{0}=-0.30 \pm 0.05$ \& $ m = 0.70 \pm 0.02 $ in the $m - q_{0}$ plane obtained by fitting derived model with $H(z)$ + SN Ia data.}
\end{figure}
The density plot of H(z) and SN Ia data are shown in Figures 4 and 5 respectively.\\
\begin{figure}[H]
\begin{tabular}{rl}
\includegraphics[width=7.5cm]{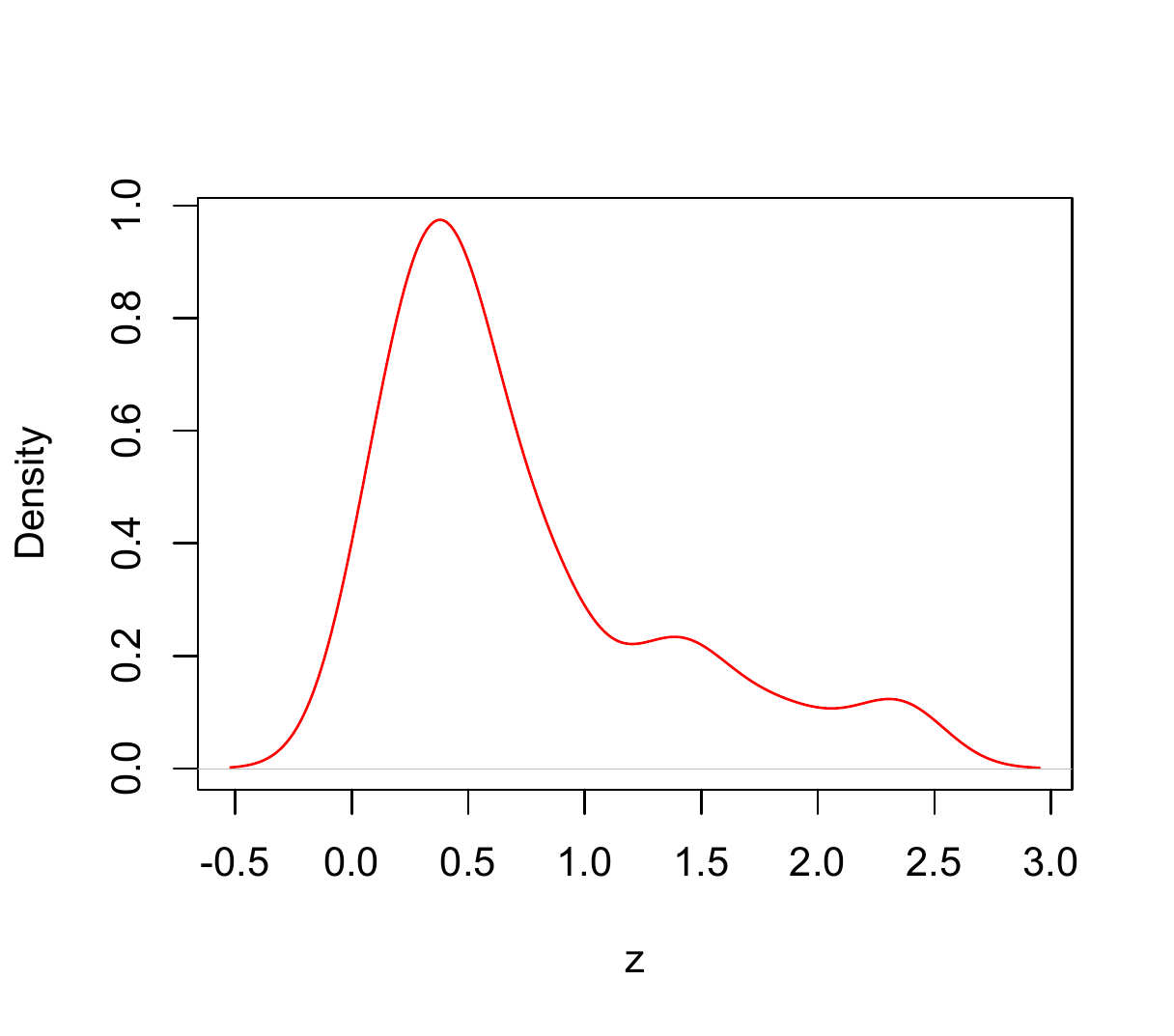}
\includegraphics[width=7.5cm]{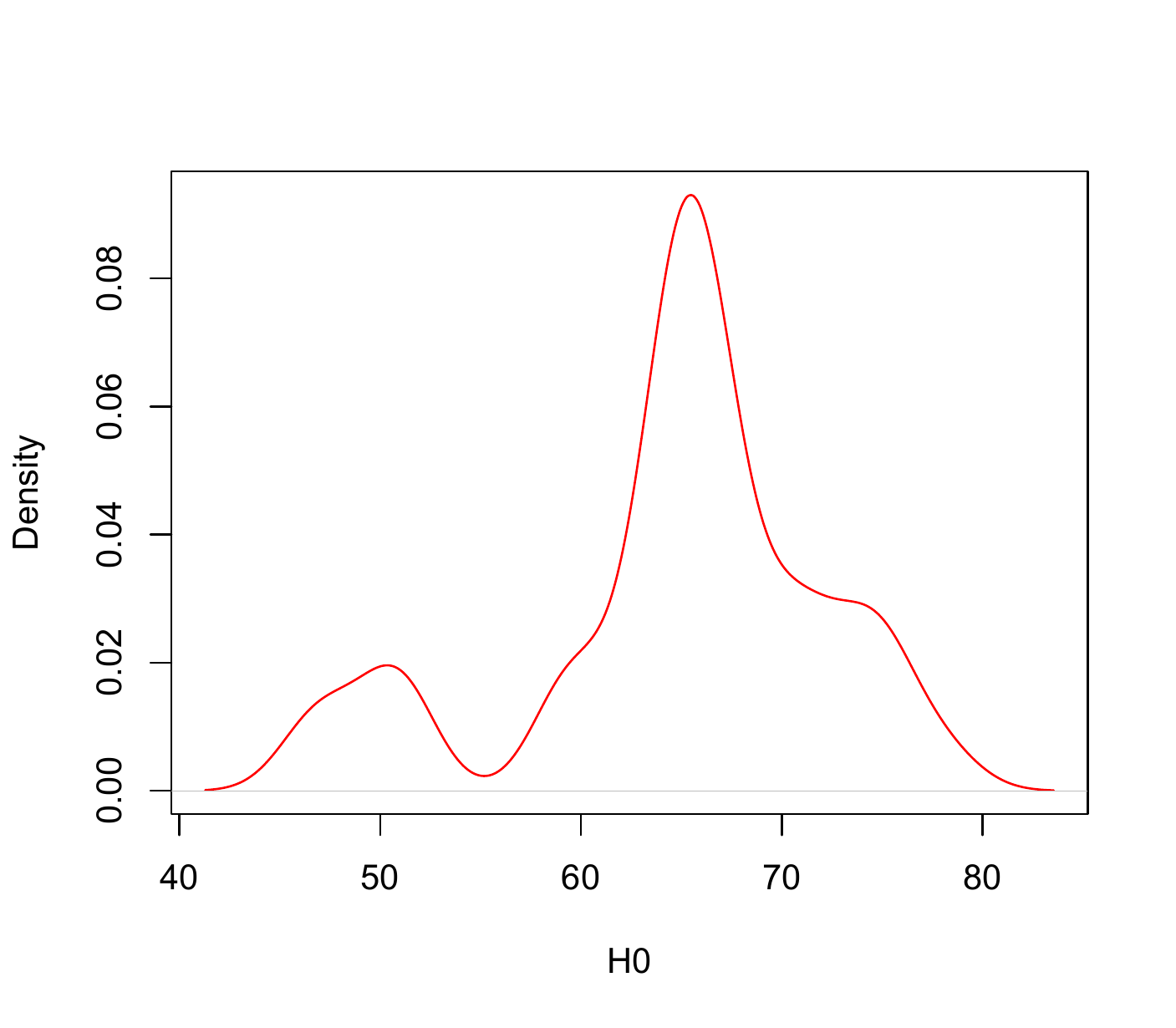}
\end{tabular}
\caption{Density plot of red-shift and Hubble's parameter of observational Hubble data points}
\end{figure}
\begin{figure}[H]
\begin{tabular}{rl}
\includegraphics[width=7.5cm]{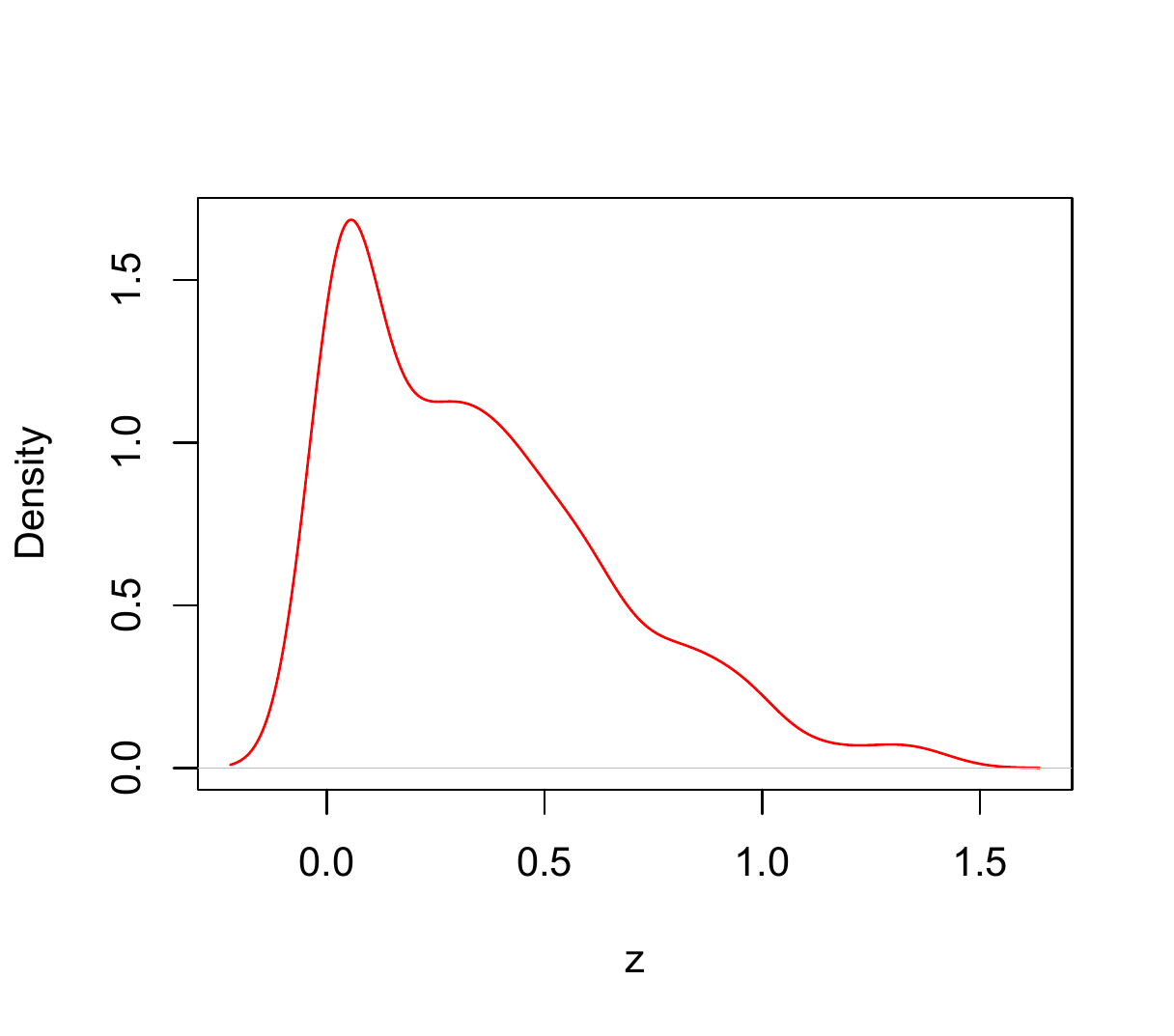}
\includegraphics[width=7.5cm]{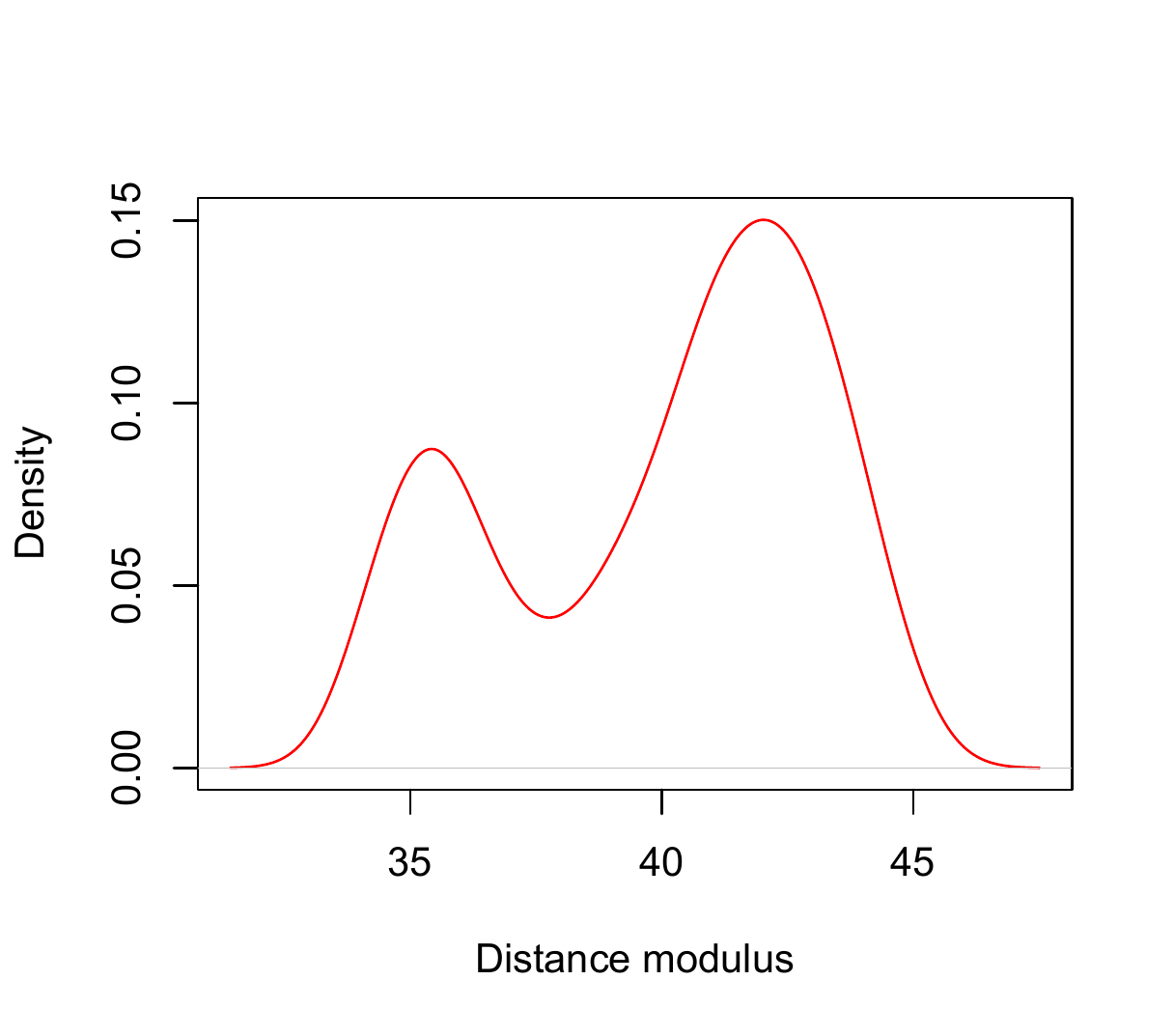}
\end{tabular}
\caption{Density plot of red-shift and distance modulus of SN Ia data}
\end{figure}
\begin{figure}[H]
\begin{center}
\includegraphics[width=9.0cm]{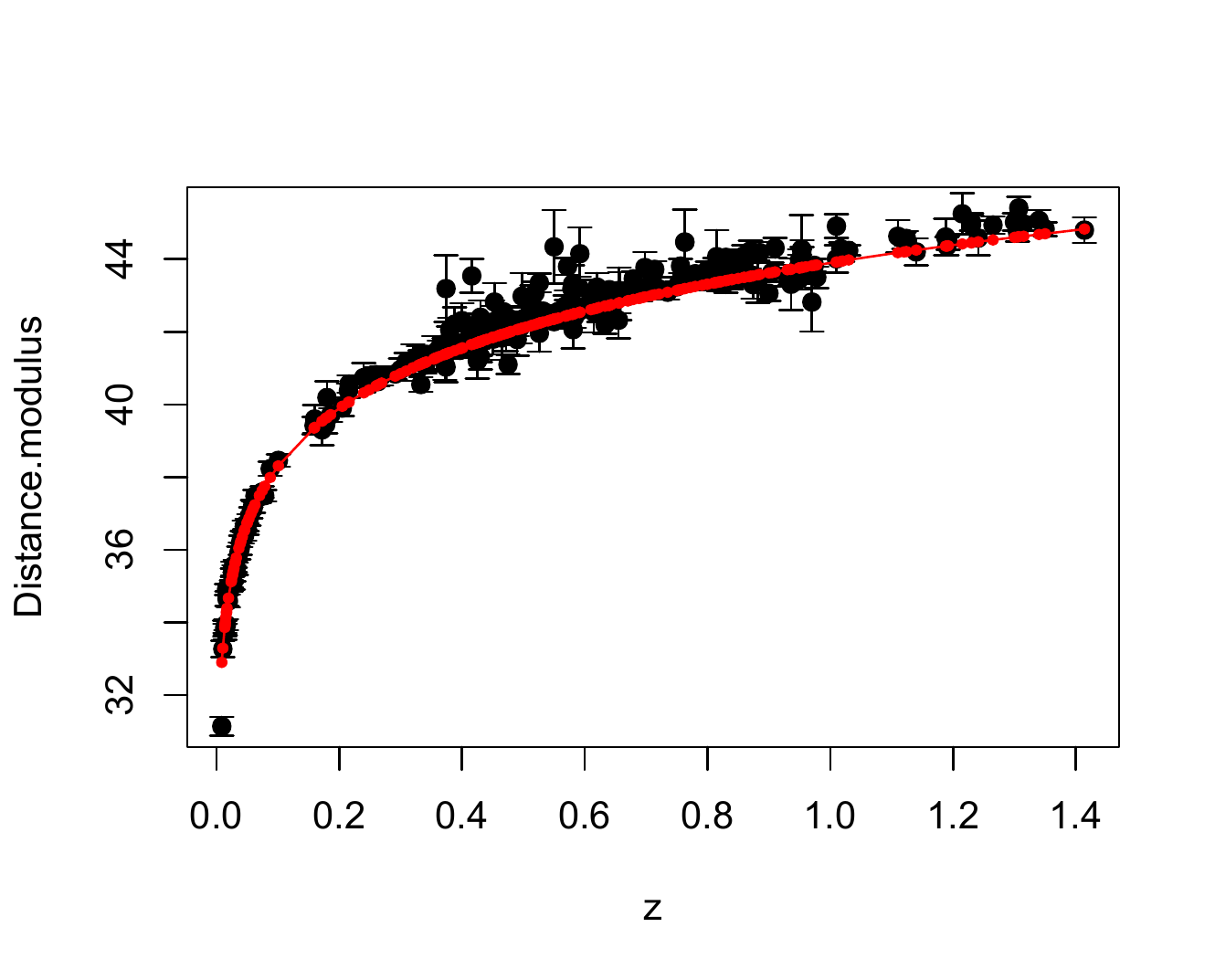}
\end{center}
\caption{The observational 580 SN Ia data points are shown with error bar (black color). The best fit model 
distance modulus $(\mu(z))$ curve (solid red points online) based on theoretical values is shown versus z.}
\end{figure}
\begin{figure}[H]
\begin{center}
\includegraphics[width=9.0cm]{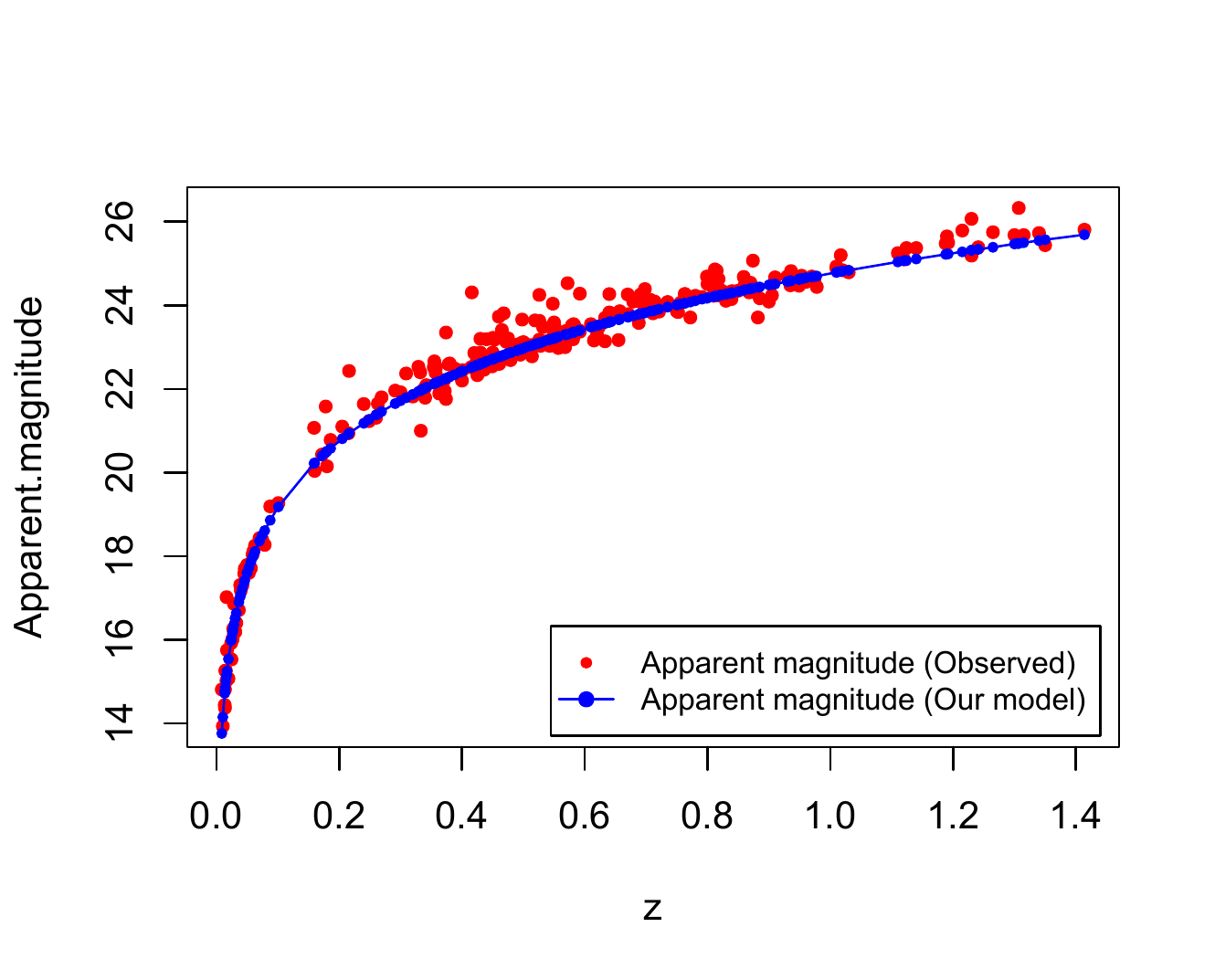}
\end{center}
\caption{Apparent magnitude versus red-shift best fit curve.}
\end{figure}
In order to get the best fit curve of derived model with 580 SN Ia data points for distance modulus and apparent magnitude, we compute $R^{2}$ test by using the following statistical formula
\begin{equation}
\label{r-mu}
R^{2}_{SN}(\mu)=1-\frac{\sum_{i}^{580}[(\mu_{i})_{obs}-(\mu_{i})_{th}]^2}{\sum_{i}^{580}[(\mu_{i})_{obs}-(\mu_{i})_{mean}]^{2}}
\end{equation}
where $(\mu_{i})_{obs}$ and $(\mu_{i})_{th}$ are the observed and theoretical values of distance modulus. $R^2 = 1$ corresponds to the ideal case when the observed data and corresponding values of theoretical function merge exactly. We find $R^{2}_{SN}(\mu) = 0.975$ and root mean square error (RMSE)= 7.17 for the model under consideration with 580 SN Ia data which reflect nice agreement of derived model with SN Ia observations. This consistency is graphed in Fig.6  in which the observational 580 SN Ia data points are shown with error bar (black color) and the best fit model distance modulus $\mu(z)$ curve (solid red points online) is shown versus z. Similarly we compute $R^{2}$ test for apparent magnitude and find that $R^{2}_{SN}(m_{b}) = 0.979$ with root mean square error (RMSE)= 6.21 which also shows appreciable consistency of derived model with observations. This result is graphed in Fig. 7. We have compiled the numerical result in Table 1.  
\begin{table*}
\small
\caption{Summary of the numerical result.\label{tbl-1}}
\begin{tabular}{@{}crrrrrrrrrrr@{}}
\hline
\hline
Source/Data~~&~~~Model parameters~~& Values at present \\
\hline
\hline
H(z) &~~~~ m ~~~ & $0.68\pm 0.04$ \\
H(z) &~~~~ $q_{0}$ ~~~ & $-0.32\pm 0.03$ \\
H(z)+ SN Ia &~~~~ m ~~~ & $0.70\pm 0.02$ \\
H(z)+ SN Ia &~~~~ $q_{0}$ ~~~ & $-0.30\pm 0.05$ \\
H(z) &~~~~ $H_{0}$ ~~~ & $65.53\pm 1.9$ \\
H(z) &~~~~ $(\Omega_{b})_{0}$ ~~~ & $0.68\pm 0.06$ \\
\hline
\end{tabular}
\end{table*}
\section{Consequences of Energy conditions}\label{section4}
Some important issues in theoretical cosmology have been discussed with help of energy
conditions (ECs) namely weak energy condition (WEC), null energy condition (NEC), dominant energy condition (DEC) and strong energy condition (SEC). In particular the violation of
SEC implies that the important problem of accelerated expansion of universe is
supported by anisotropic universe \cite{Visser/2000}. The ECs in $f(R,T)$ gravity
theory by incorporating conservation of energy-momentum tensor is presented in the
literature \cite{Chakraborty/2013} and it has been analyzed that $T$ sector can not be chosen arbitrarily but it has special form. The ECs in modified gravitational field equations are given as follows \cite{Moraes/2017}
\begin{equation}\label{ec}
\begin{split}
WEC \Longleftrightarrow \rho \geq 0; \rho + \bar{p} \geq 0, \\
NEC \Longleftrightarrow \rho + \bar{p} \geq 0, \\
DEC \Longleftrightarrow \rho \geq |\bar{p}|, \\
SEC \Longleftrightarrow \rho +3 \bar{p} \geq 0. \\
\end{split}
\end{equation}
The behaviors of the above ECs are depicted in Fig. 10 with $m=0.7 ~\&~ \gamma=0.5$. From Fig. 10, we observe that all the energy conditions are violated for the fixed values of the free parameters. The values of free parameters are obtained by bounding the derived model with observational data and positive energy density.
\begin{figure}[H]
\begin{tabular}{rl}
\includegraphics[width=7.5cm]{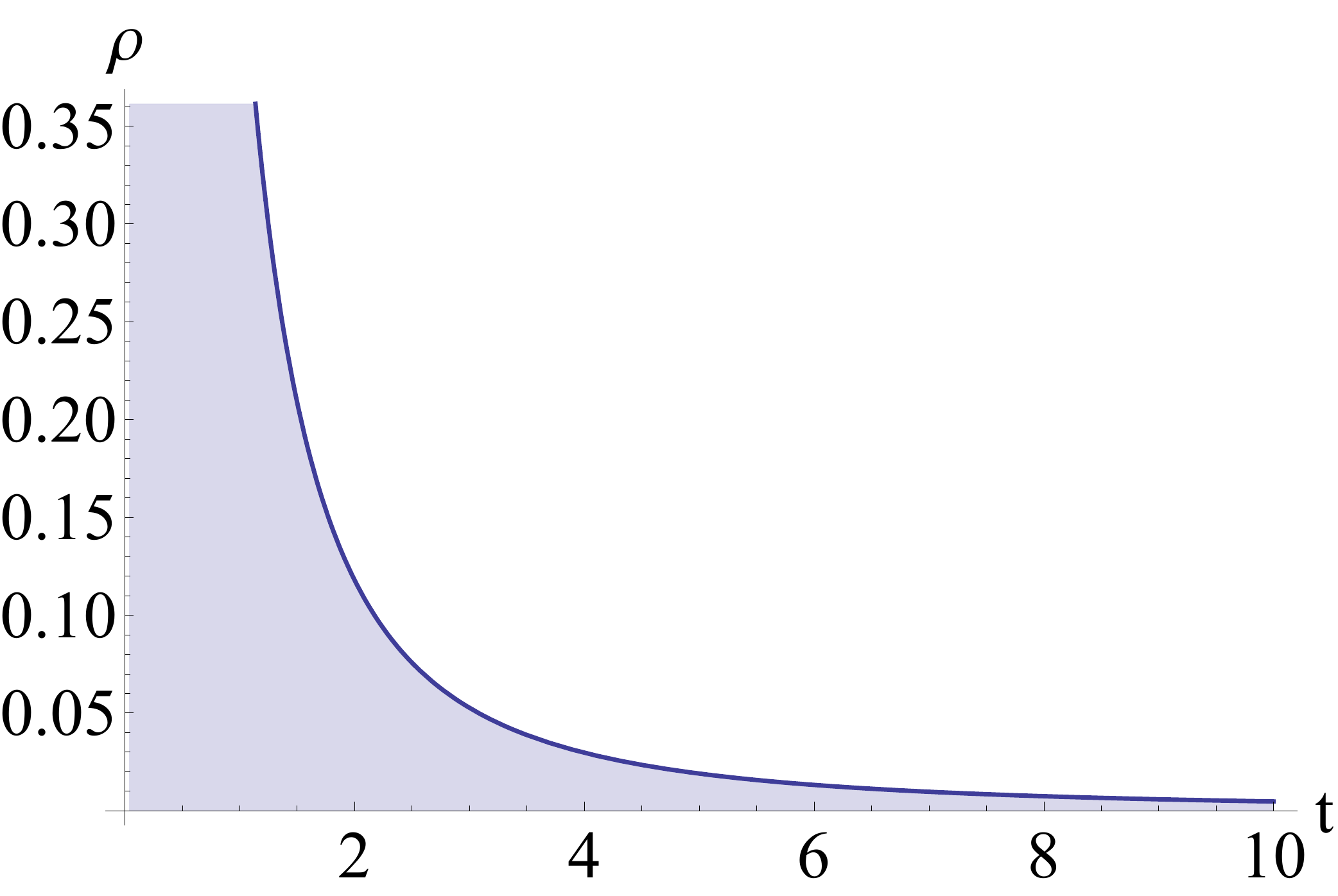}
\includegraphics[width=7.5cm]{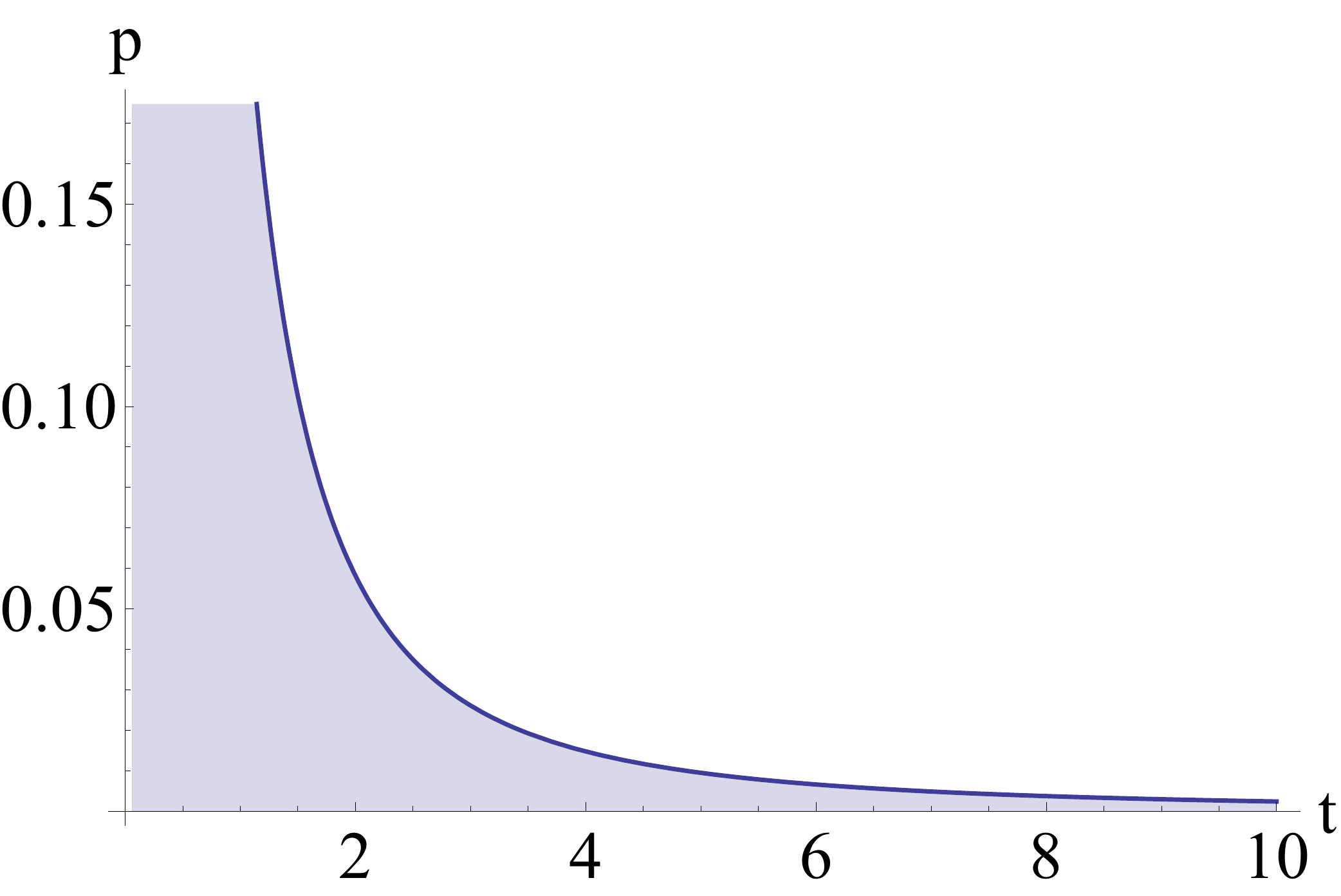}
\end{tabular}
\caption{Energy density and pressure vs. t with $\gamma = 0.5$ \& $m = .70$.}
\end{figure}
\begin{figure}[H]
\begin{center}
\includegraphics[width=9.0cm]{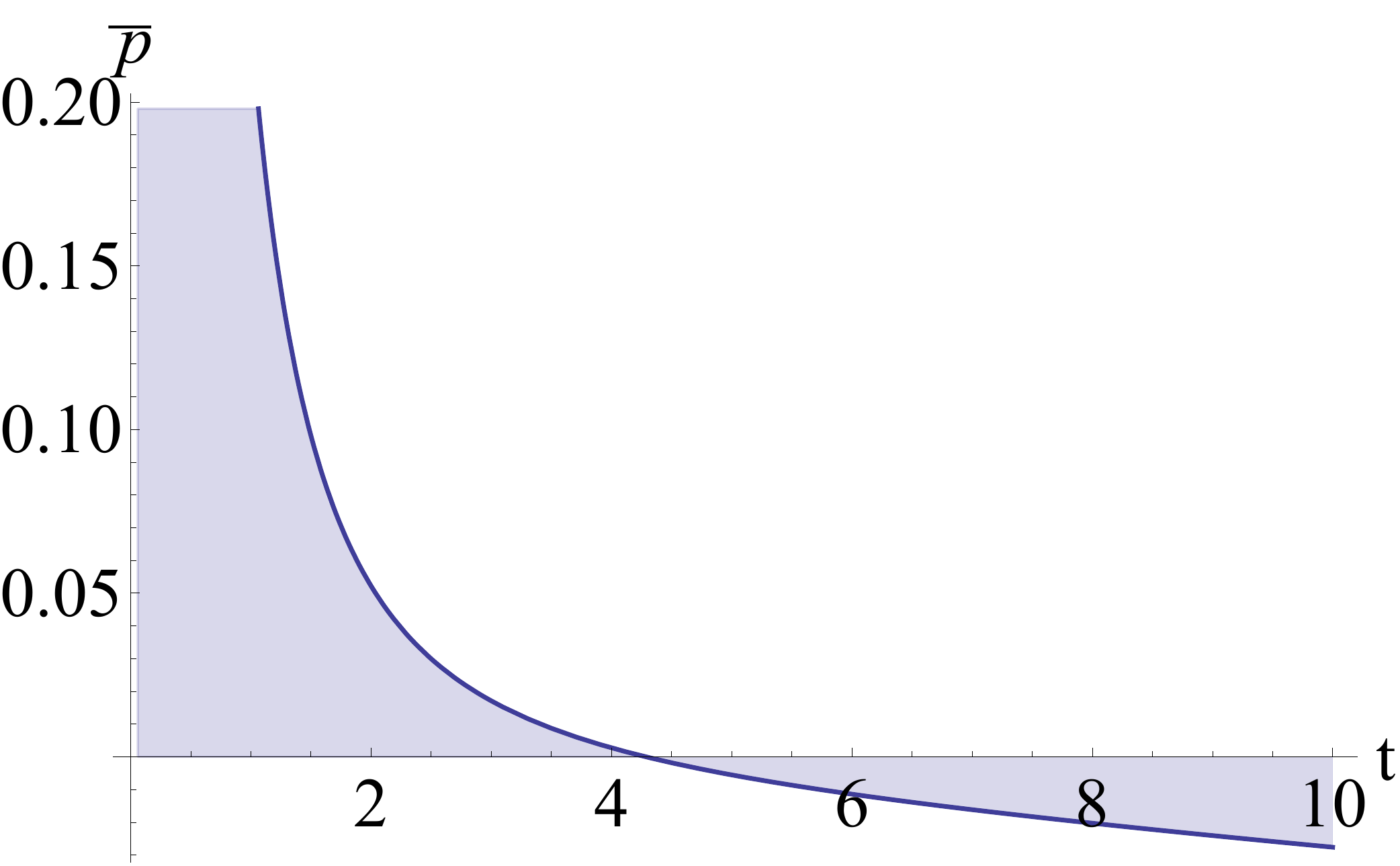}
\end{center}
\caption{Bulk viscous pressure vs. t with $\gamma = 0.5$ \& $m = .70$.}
\end{figure}
\begin{figure}[H]
\begin{center}
\includegraphics[width=9.0cm]{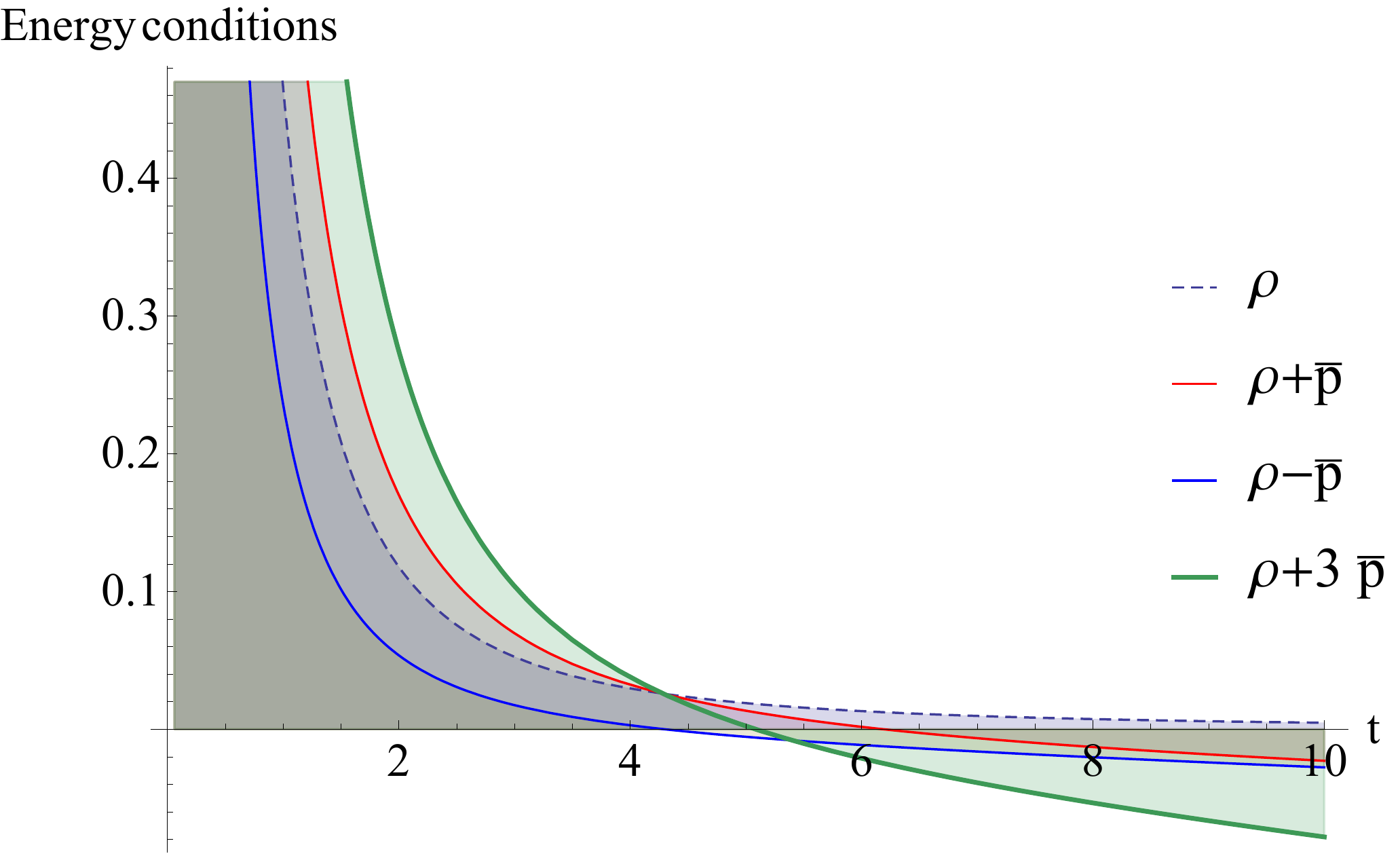}
\end{center}
\caption{Energy conditions vs. t with $\gamma = 0.5$ \& $m = .70$.}
\end{figure}
\section{Om(z) diagnostic analysis}\label{section5}
In literature, the state finder parameters $r-s$ and analysis of Om diagnostic
are used to study dark energy models \cite{Sahni/2008}. The Om(z), is a combination of the Hubble parameter $H$ and the cosmological redshift $z$. The Om(z) parameter in modified gravity is given as \cite{Sahoo/2018}
\begin{equation}
Om(z)=\frac{\left[\frac{H(z)}{H_0}\right]^2-1}{(1+z)^3-1}
\end{equation}
where $H_0$ is the present Hubble parameter. The negative, zero and positive values of Om$(z)$ represents the quintessence ($\omega>-1$), $\Lambda$CDM  and phantom ($\omega<-1$) DE models respectively \cite{Shahalam/2015}. In the present model the Om(z) parameter is obtained as
\begin{equation}
Om(z)= \frac{(1+z)^{2m}-1}{(1+z)^3 - 1}
\end{equation}
and its behaviour can be seen from \textbf{Fig. 11}.
\begin{figure}[H]
\begin{center}
\includegraphics[width=9.0cm]{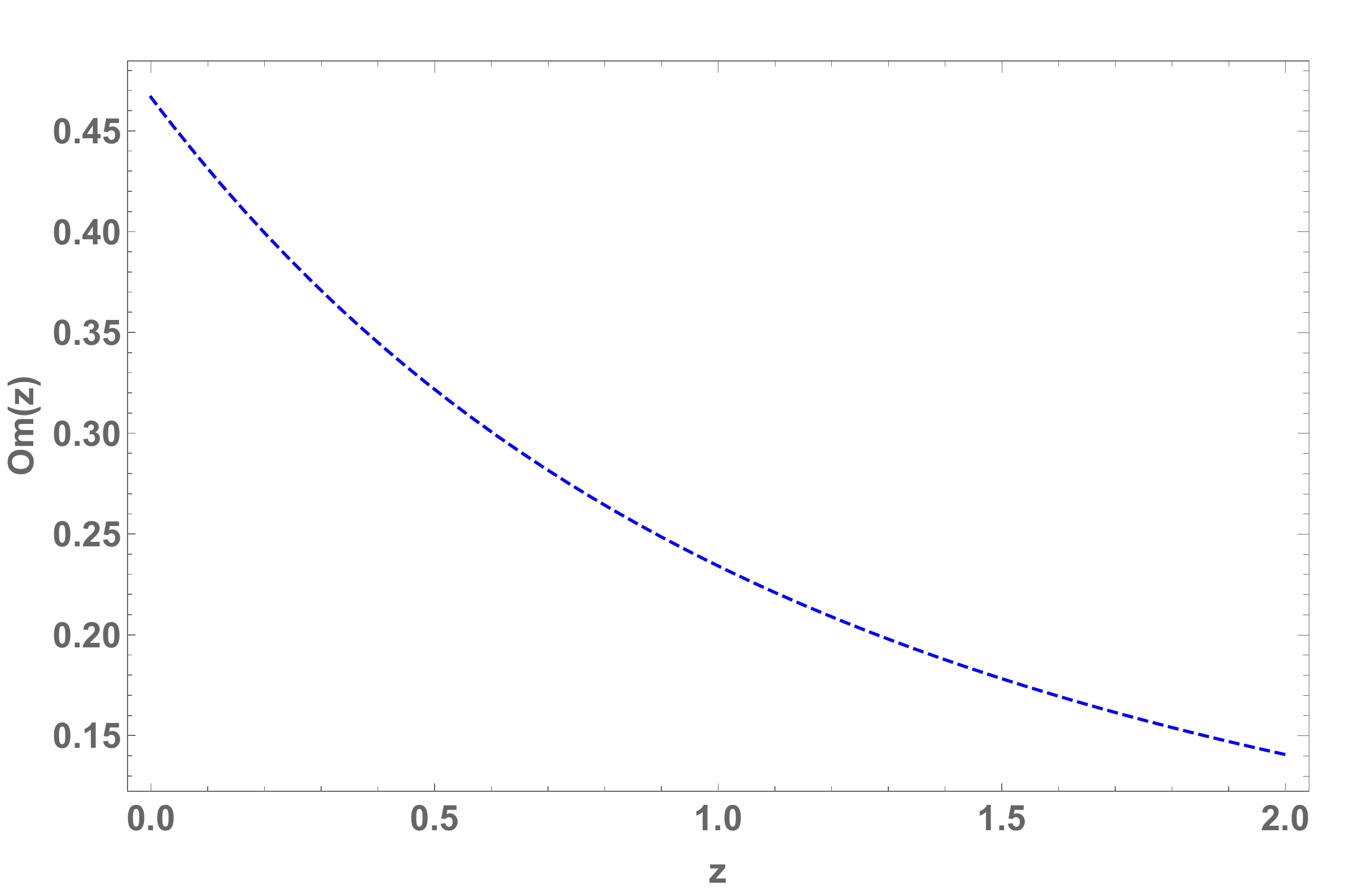}
\end{center}
\caption{Variation of Om(z) against $z$ with $m = .70$.}
\end{figure}
\section{Concluding Remarks}\label{section6}
In this paper, we have estimated the numerical values of model parameters of derived model by bounding it with observational data. The numerical results are tabulated in Table I.\\
Some key observations of present study are as follows:\\
\textbullet~ The computed value of deceleration parameter for derived model is $q_{0}$ ~~~= $-0.30\pm 0.05$ at 1$\sigma$ confidence level.\\
\textbullet~ The distance modulus $(\mu(z))$ and apparent distance $(m_{b}(z))$ of derived best fit model fit well to the observational data points from astronomical observations (see Fig. 6 \& 7).\\
\textbullet~ The derived model is LRS BI anisotropic universe which tends to isotropic at $t\rightarrow \infty$ $i.e.$  the anisotropy is null for larger time.\\
\textbullet~ In the derived model, we find $(\Omega_{b})_{0} = 0.68\pm 0.06$ at $1\sigma$ level of 36 OHD points. It declares the significant contribution of bulk viscosity in the present universe.\\
\textbullet~ The NEC, WEC, DEC and SEC are violated in our present model (Fig. 10). It is worth to mention here that due to the current accelerated expansion of the universe, SEC must be violated \cite{Barcelo/2002}. Hence, we can conclude that our derived model is a spar with the current accelerated expansion of the universe.\\  
\textbullet~ The Om(z) is plotted with respect to redshift in the range $0\leq z \leq 2$ in Fig. 11. The value of Om(z) is decreasing as the redshift parameter increasing in the above range. The positive value of Om(z) indicates that the model is in phantom era.
\section*{ACKNOWLEDGMENT}
PKS acknowledges CSIR, New Delhi, India for financial support to carry out the Research project [No.03(1454)/19/EMR-II Dt.02/08/2019]. We are very much grateful to the honourable referee and the editor for the illuminating suggestions that have significantly improved our work in terms of research quality and presentation. 

\end{document}